\documentclass[aps,pre,twocolumn,groupedaddress,amsmath,showpacs]{revtex4}
	\usepackage{bm}
	\usepackage{graphicx}
  \usepackage{monops} 
\newcommand\beq{\begin{equation}}
\newcommand\eeq{\end{equation}}
\newcommand\beqa{\begin{eqnarray}}
\newcommand\eeqa{\end{eqnarray}}

\begin{document} 
\title{Comments on nonlinear viscosity and Grad's moment method } 
\author{Andr\'es Santos}
\email[]{andres@unex.es}
\homepage{http://www.unex.es/fisteor/andres}
\affiliation{Departamento de F\'{\i}sica, Universidad de Extremadura, 
Badajoz, E-06071, Spain}
\date{\today}

\begin{abstract}
It is shown that  the steady unidirectional flow with vanishing heat 
flux considered by B. C. Eu [Phys. Rev. E \textbf{65}, 
031202 (2002)], and earlier by  Uribe and Garc\'{\i}a--Col\'{\i}n [Phys. 
Rev. E \textbf{60}, 4052 (1999)], is inconsistent with the exact 
conservation laws of mass, momentum, and energy. The inconsistency does not 
lie in the assumed symmetry properties of the flow but in the stationarity 
assumption.
The unsteady problem is considered and its solution from the Boltzmann equation for Maxwell molecules is given.
\end{abstract}

\maketitle

In a recent paper \cite{E02}, Eu analyzed a \textit{steady} unidirectional 
flow at uniform temperature and derived the equations for the stress tensor 
elements from the Boltzmann equation by means of Grad's moment method. 
These are essentially the same state and the same method as those considered 
by Uribe and Garc\'{\i}a--Col\'{\i}n \cite{UGC99}, except that the 
transversal velocity gradients were assumed to vanish in Ref.~\cite{UGC99}, 
while they are included in the analysis of Ref.~\cite{E02}. The major aim of 
this Comment is to show that the steady unidirectional flow at uniform 
temperature of Refs.\ \cite{E02,UGC99} is inconsistent with
the exact macroscopic conservation equations.

For a dilute monatomic gas, the macroscopic balance equations expressing the 
conservation of mass, momentum, and energy are \cite{GM84,CC70}
\beq
D_t n+n\nabla\cdot \mathbf{u}=0,
\label{1}
\eeq
\beq
D_t \mathbf{u}+\frac{1}{mn}\nabla\cdot\mathsf{P}=\mathbf{0},
\label{2}
\eeq
\beq
D_t T+\frac{2}{3nk_B} \left(\nabla \cdot \mathbf{q}+\mathsf{P}:\nabla 
\mathbf{u}\right)=0,
\label{3}
\eeq
where $D_t\equiv \partial_t+\mathbf{u}\cdot \nabla$ is the material time 
derivative, $n$ is the local number density, $\mathbf{u}$ is the local flow 
velocity, $T$ is the local temperature, $m$ is the mass of a particle, $k_B$ 
is the Boltzmann constant, 
$\mathbf{q}$ is the heat flux vector, and $\mathsf{P}$ is the pressure (or stress) 
tensor.
The flow considered in Refs.\ \cite{E02,UGC99} is characterized by the 
following properties (not necessarily independent): (a) it is a 
unidirectional flow, i.e., 
$\mathbf{u}(\mathbf{r})=u_x(\mathbf{r})\widehat{\mathbf{x}}$, where 
$\widehat{\mathbf{x}}$ is the unit vector along the flow direction; (b) the 
temperature is uniform, $\nabla T=\mathbf{0}$; (c) the heat flux vanishes, 
$\mathbf{q}=\mathbf{0}$; (d) the pressure tensor is uniform; and (e) the 
state is stationary, i.e., $\partial_t\to 0$.
Let me consider first the geometrical properties (a)--(d) separately from 
the stationarity assumption (e).
Application of assumptions (a)--(d) on the exact balance equations 
(\ref{1})--(\ref{3}) yields
\beq
\partial_t n+\frac{\partial}{\partial x}\left(n u_x\right)=0,
\label{4}
\eeq
\beq
\partial_t u_x+u_x\frac{\partial}{\partial x}u_x=0,
\label{5}
\eeq
\beq
\frac{3nk_B}{2}\partial_t T+P_{xx}\frac{\partial u_x}{\partial 
x}+P_{xy}\frac{\partial u_x}{\partial y}+P_{xz}\frac{\partial u_x}{\partial 
z}=0.
\label{6}
\eeq
Are Eqs.~(\ref{4})--(\ref{6}) consistent with a steady state?
Equation (\ref{5}) shows that the flow velocity is stationary if and only if 
the flow is incompressible, i.e., if $\partial u_x/\partial x=0$. In that 
particular case, Eq.~(\ref{4}) is consistent with a stationary density if 
and only if the density is uniform as well. But, even if $\partial 
u_x/\partial x=0$ and $n=\text{const}$,  energy balance 
equation (\ref{6}) implies that the temperature cannot be stationary but 
monotonically increases with time due to viscous heating effects (note that 
$P_{xy}\partial u_x/\partial y <0$ and $P_{xz}\partial u_x/\partial z <0$ 
because of physical reasons). Therefore, the steady state assumption (e) is 
incompatible with assumptions (a)--(d), except in the trivial case 
$\nabla u_x=\mathbf{0}$, i.e., at equilibrium.
The \textit{unsteady} Boltzmann equation for the incompressible 
unidirectional flow with $\partial u_x/\partial x=\partial u_x/\partial 
z=0$, $\partial u_x/\partial y=\text{const}$, usually referred to as uniform 
shear flow or homoenergetic simple shear flow, has been solved \textit{exactly} for arbitrary shear rates in 
the case of Maxwell molecules \cite{IT56,TM80,SG95}. An analogous solution 
has been obtained in the case of the BGK model kinetic equation for more 
general interactions \cite{Z79,SB91,GS95}.

In the case of a compressible flow in the absence of transversal gradients, i.e., $\nabla u_x \| 
\widehat{\textbf{x}}$, Eq.~(\ref{5}) 
shows that the flow velocity is 
necessarily unsteady. According to  continuity equation (\ref{4}), it 
is still mathematically possible that $\partial_t n=0$ if the product $nu_x$ 
is uniform, i.e., $n(x)u_x(x,t)=K(t)$. Insertion of this condition into 
Eq.~(\ref{5}) yields $K^{-2}\dot{K}(t)=n^{-2}n'(x)=-A^{-1}$, where the dot 
denotes a time derivative, the prime denotes a spatial derivative, and $A$ 
is a constant.
The solution to these equations is simply $n(x)=A/(x-x_0)$, 
$u_x(x,t)=a(x-x_0)/(1+at)$, where $x_0$ and $a$ are constants. This 
mathematical solution is unphysical unless the problem is restricted to 
the half domain $x>x_0$ ($x<x_0$) if $A>0$ ($A<0$). But even in that case 
the existence of a nonuniform density is in conflict with the 
uniformity assumptions (b) and (d) because in a dilute gas $n=p/k_BT$, where 
$p=\frac{1}{3}\text{tr}\, \mathsf{P}$ is the hydrostatic pressure. In 
summary, assumptions (a)--(d) do not contradict the conservation laws 
(\ref{1})--(\ref{3}) in the compressible flow with  $\partial u_x/\partial 
x\neq 0$ if and only if the three hydrodynamic quantities (density, flow 
velocity, and temperature) are unsteady, so  assumption (e) is again 
incompatible with (a)--(d).

Strictly speaking,  assumptions (b) and (d), i.e.,  uniform temperature 
 and pressure tensor, were not explicitly stated in Ref.~\cite{UGC99}. 
 Actually, assumption (b) is implicit in (c); otherwise, one would have a 
thermal gradient that does not produce any heat flux, what is at odds with 
the second principle of thermodynamics.
As for assumption (d), it was replaced in Ref.~\cite{UGC99} by a weaker one: 
(d') the \textit{irreversible} part of the pressure tensor is uniform, 
namely $\partial(P_{xx}-p)/\partial x=0$.
{}From a physical point of view it seems difficult to reconcile a uniform normal stress difference $P_{xx}-p$ with a nonuniform hydrostatic pressure $p$. In any case,
let me drop  conditions (b) and (d) for the moment and prove that conditions (a), (c), (d'), and (e) are 
also inconsistent with the conservation laws, except at equilibrium.
Application of (a), (c), and (e) on Eqs.~(\ref{1})--(\ref{3}) gives
\beq
\frac{\partial}{\partial x}\left(nu_x\right)=0,
\label{n1}
\eeq
\beq
mnu_x  \frac{\partial}{\partial x} u_x+\frac{\partial}{\partial x} P_{xx}=0,
\label{n2}
\eeq
\beq
\frac{3}{2}nk_B u_x\frac{\partial}{\partial x} T+P_{xx}
\frac{\partial}{\partial x} u_x=0,
\label{n3}
\eeq
where, as in Ref.~\cite{UGC99}, the case $\nabla u_x\| \widehat{\mathbf{x}}$ 
has been considered.
Equations (\ref{n1})--(\ref{n3}) can be easily integrated to get
\beq
nu_x=n_0u_0=\text{const},
\label{n4}
\eeq
\beq
P_{xx}+mnu_x^2=P_0+mn_0u_0^2=\text{const},
\label{n5}
\eeq
\beqa
\left(\frac{3}{2}p+P_{xx}+\frac{m}{2}nu_x^2\right)u_x&=&
\left(\frac{3}{2}p_0+P_0+\frac{m}{2}n_0u_0^2\right)u_0
\nonumber\\
&=&\text{const},
\label{n6}
\eeqa
where the subscript 0 denotes quantities evaluated at some reference point 
$x=x_0$.
So far, assumption (d') has not beeen used. This condition implies that 
$P_{yy}-p=P_0-p_0=\text{const}$, so that, according to Eq.~(\ref{n5}),
\beq
p+mnu_x^2=p_0+mn_0u_0^2=\text{const}.
\label{n7}
\eeq
Insertion of Eqs.~(\ref{n5}) and (\ref{n7}) into Eq.~(\ref{n6}) yields
\beq
\left[\frac{3}{2}p_0+P_0+mn_0u_0\left(\frac{1}{2}u_0-2u_x\right)\right]\left(u_x 
-u_0\right)=0.
\label{n8}
\eeq
Both solutions of this quadratic equation are constants (but the physical 
one is $u_x=u_0$). This closes the proof that assumptions (a), (c), (d'), 
and (e) are not consistent with the conservation equations except in the 
trivial case $u_x=\text{const}$.

On the other hand, there is nothing wrong with assumptions (a)--(d) in 
the unsteady case.
What is then the right form of $n$ and $u_x$ if $\nabla u_x \| 
\widehat{\textbf{x}}$?
Since assumptions (b) and (d) imply that $n$ is 
uniform, Eq.~(\ref{4}) states that $-\partial u_x/\partial 
x=n^{-1}\dot{n}=-K(t)$, so $u_x(x,t)=K(t)(x-x_0)$. Substitution into 
Eq.~(\ref{5}) gives $K^{-2}\dot{K}=-1$, so we finally have 
$n(t)=n_0/(1+at)$, $u_x(x,t)=a(x-x_0)/(1+at)$. This simple flow is known as 
homoenergetic extension \cite{TM80,G95}. Again, the \textit{unsteady} Boltzmann 
equation can be solved \textit{exactly} for arbitrary values of the 
constant control parameter $a$ in the case of Maxwell molecules 
\cite{TM80,G95,S00}, as well as with Grad's method \cite{KDN97} and in the 
case of the BGK model kinetic equation \cite{S00} for more general 
interactions.
A situation where a transversal velocity gradient $\partial u_y/\partial x$ coexists with a longitudinal one $\partial u_x/\partial x$
has been studied by Galkin \cite{G95}.

It is illustrative to recall the application of Grad's method to the 
\textit{unsteady} unidirectional flow at uniform temperature with no 
transversal gradients \cite{KDN97}. We have seen above that conservation of 
mass and momentum imply that $n(t)=n_0/(1+at)$ and $\gamma_x(t)\equiv 
\partial u_x(x,t)/\partial x=a n(t)/n_0$. 
Positive values of the longitudinal deformation rate $\gamma_x$ represent 
some sort of ``explosion'' (or expansion) flow, while negative values 
represent an ``implosion'' (or condensation) flow \cite{UGC99}.
The energy balance equation (\ref{6}) becomes
\beq
\partial_t p +\left(p+\frac{2}{3}P_{xx}\right)\gamma_x=0.
\label{n9}
\eeq
In Grad's method, this equation is coupled to the (approximate) evolution equation for the 
stress element $P_{xx}$ \cite{KDN97}:
\beq
\partial_t 
\left(P_{xx}-p\right)+\left(P_{xx}-p+\frac{4}{3}P_{xx}\right)\gamma_x=-\mu 
\left(P_{xx}-p\right),
\label{n10}
\eeq
where nonlinear terms have been neglected on the 
right-hand side and $\mu=p/\eta_{\text{NS}}$ is 
 an effective collision frequency, $\eta_{\text{NS}}$ being the 
Navier--Stokes shear viscosity.
Equation (\ref{n10}) is equivalent to Eq.~(21) of Ref.~\cite{UGC99}, except 
that the time derivative is absent in the latter.
Without  the time derivative operator, however, Eq.~(\ref{n10}) cannot be 
made consistent with Eq.~(\ref{n9}). 
After a transient regime, the system reaches a generalized hydrodynamic 
regime with $P_{ij}(t)=p(t)P_{ij}^*(\gamma_x^*(t))$, where 
$\gamma_x^*(t)\equiv \gamma_x(t)/\mu(t) $ is the 
\textit{reduced} longitudinal deformation rate. In general, from Eqs.~(\ref{n9}) and 
(\ref{n10}) one gets a nonlinear first-order ordinary differential equation 
for $P_{xx}^*(\gamma_x^*)$ \cite{S00,KDN97}. 
In the special case of Maxwell molecules (i.e $\mu\propto n$), $\gamma_x^*$ (but not $\gamma_x$) is independent of time 
 so one gets from Eqs.~(\ref{n9}) and (\ref{n10}) an 
algebraic quadratic equation whose physical solution is $P_{xx}^*=3-2P_{yy}^*$ with
\beq
P_{yy}^*(\gamma_x^*)=\frac{3}{8\gamma_x^*}\left(\sqrt{1+\frac{4}{3} 
\gamma_x^*+4{\gamma_x^*}^2}-1+2\gamma_x^*\right).
\label{n11}
\eeq
It turns out that this result for Maxwell molecules goes beyond the 
scope of Grad's method since it can be exactly derived from the Boltzmann 
equation \cite{TM80,G95,S00}.

To the best of my knowledge, Grad's method has not been applied yet to the unsteady unidirectional flow at uniform temperature with both longitudinal and transversal gradients. For this flow, an analysis of Eqs.~(\ref{4}) and (\ref{5}) shows that, in addition to  $n(t)=n_0/(1+at)$ and $\gamma_x\equiv \partial u_x/\partial x=an(t)/n_0$, one generally has shear rates $\gamma_{yx}\equiv \partial u_x/\partial y=a_2n(t)/n_0$ and $\gamma_{zx}\equiv \partial u_x/\partial z=a_3n(t)/n_0$, where $a$, $a_2$, and $a_3$ are independent constants. Standard application of Grad's method yields the following evolution equation for the elements of the stress tensor:
\beq
\partial_t P_{ij}+P_{ij}\frac{\partial u_x}{\partial x}+\sum_{k=1}^3\left(P_{ik}\delta_{jx}+P_{jk}\delta_{ix}\right)\frac{\partial u_x}{\partial x_k}=-\mu(P_{ij}-p\delta_{ij}).
\label{n12}
\eeq
Taking the trace in this equation one recovers the energy conservation equation (\ref{6}). 
It is worth noting that Eq.~(\ref{n12}) can again be obtained exactly from the Boltzmann equation  in the case of Maxwell molecules. For this interaction potential, the reduced rates $\gamma_x^*\equiv \gamma_x/\mu$, $\gamma_{yx}^*\equiv \gamma_{yx}/\mu$, and $\gamma_{zx}^*\equiv \gamma_{zx}/\mu$ are constants and Eq.~(\ref{n12}) yields a coupled set of algebraic equations for $P_{ij}^*\equiv P_{ij}/p$. Without loss of generality \cite{note1} we can choose $\gamma_{zx}^*=\gamma_{yx}^*$, so that $P_{yy}^*=P_{zz}^*$ and $P_{xy}^*=P_{xz}^*$. In that case, Eq.~(\ref{n12}) yields
\beq
\gamma_x^*=\frac{3P_{yy}^*\left(1-P_{yy}^*\right)-4{P_{xy}^*}^2}{2P_{yy}^*\left[2{P_{xy}^*}^2-P_{yy}^*\left(3-2P_{yy}^*\right)\right]},
\label{n13}
\eeq
\beq
\gamma_{yx}^*=\frac{P_{xy}^*\left(3-P_{yy}^*\right)}{2P_{yy}^*\left[2{P_{xy}^*}^2-P_{yy}^*\left(3-2P_{yy}^*\right)\right]}.
\label{n14}
\eeq
These two equations include as particular cases the homoenergetic extension flow ($\gamma_{yx}^*=0$), in which case Eq.~(\ref{n11}) is recovered, as well as the uniform shear flow ($\gamma_{x}^*=0$), where ${P_{xy}^*}^2=\frac{3}{4}P_{yy}^*\left(1-P_{yy}^*\right)$ and $P_{yy}^*$ is the solution of the cubic equation $4{\gamma_{yx}^*}^2{P_{yy}^*}^3=3\left(1-P_{yy}^*\right)$.

\begin{figure}[tbp]
\includegraphics[width=0.9\columnwidth]
{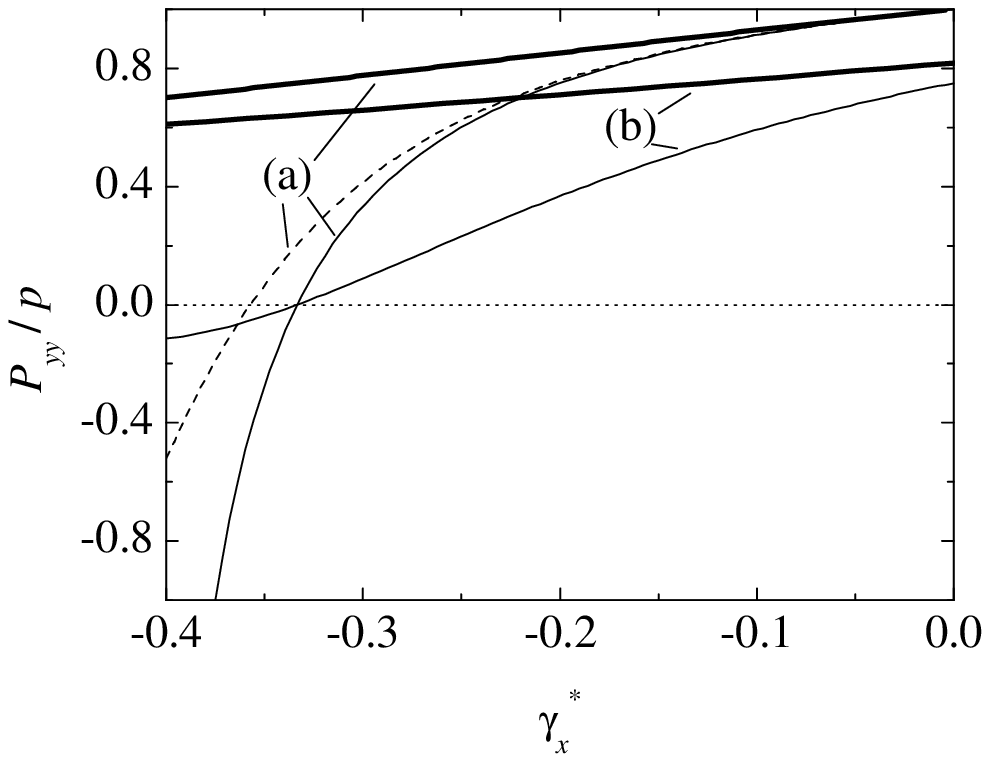}
 \caption{Plot of the normal pressure element $P_{yy}$ relative to the 
 hydrostatic pressure $p$, as derived in Refs.~\protect\cite{E02,UGC99}, versus the longitudinal deformation rate in the range  
 $-0.4\leq \gamma_x^*\leq 0$ for (a) zero shear rate 
 ($\gamma_{yx}^*=0$) and (b) $\gamma_{yx}^*=\frac{1}{2}$.
 The thin solid lines correspond to the linear approximation, 
 Eq.~(\protect\ref{7}), while the dashed line in case (a) corresponds to the 
 nonlinear approximation, Eq.~(\protect\ref{7.2}). 
 The thick solid lines represent the exact 
 results for Maxwell molecules.\label{fig1}}
\end{figure}

The inconsistency of assumption (e) on the stationarity of the flow 
geometrically characterized by (a)--(d) [or (a), (c), and (d')] manifests itself in the results 
obtained in Refs.~\cite{E02} and \cite{UGC99} for the stress tensor 
elements from Grad's method. Assuming cylindrical symmetry 
($P_{xy}=P_{xz}$, $P_{yy}=P_{zz}$), Eu gets the following expression for the 
normal pressure element $P_{yy}$ (neglecting nonlinear terms in the 
collisional integrals) \cite{E02}:
\beq
P_{yy}^*=3\frac{(1+2\gamma_x^*)(1+3{\gamma_x^*})}{(1+2\gamma_x^*)(3+7{\gamma_x^*})
+4 {\gamma_{yx}^*}^2}.
\label{7}
\eeq
 According to 
Eq.~(\ref{7}), $P_{yy}$ becomes \textit{negative} at least in the interval 
$-\frac{3}{7}<\gamma_x^*<-\frac{1}{3}$, regardless of the value of 
$\gamma_{yx}^*$. This is unphysical because the diagonal elements of the 
pressure tensor are positive definite quantities. 
Setting $\gamma_{yx}^*=0$, Eq.~(\ref{7}) reduces to the result derived by 
Uribe and Garc\'{\i}a--Col\'{\i}n in the linear approximation \cite{UGC99}. 
When the nonlinear terms are included, they get \cite{UGC99}
\beq
P_{yy}^*=8+\frac{49}{3}\gamma_x^*-7\sqrt{1+\frac{94}{21}\gamma_x^*+ 
\frac{49}{9}{\gamma_x^*}^2}.
\label{7.2}
\eeq
According to this expression, $P_{yy}<0$ if $\gamma_x^*<-\frac{5}{14}$.
The prediction of 
negative values of $P_{yy}$ can be observed in Fig.~\ref{fig1}, which shows 
the ratio $P_{yy}/p$ given by Eqs.~(\ref{7})  and (\ref{7.2}) in the range 
$-0.4\leq \gamma_x^*\leq 0$ for vanishing shear rate ($\gamma_{yx}^*=0$) and 
for $\gamma_{yx}^*=\frac{1}{2}$. 
Comparison with the exact results (\ref{n11}) for $\gamma_{yx}^*=0$ and  Eqs.\ (\ref{n13}) and (\ref{n14}) for $\gamma_{yx}^*\neq 0$ in the case of Maxwell molecules  shows that the predictions (\ref{7}) and (\ref{7.2}) 
are only valid in the Navier-Stokes domain of small gradients, where 
$P_{yy}/p\approx 1+\frac{2}{3}\gamma_x^*$.
It is worth noting that the application to the unidirectional flow of a 
rheological theory by Eu \cite{E02b} also yields unphysical negative values 
for the diagonal elements of the pressure tensor \cite{UGC02}.

In Ref.~\cite{E02}, Eu claims that Grad's 
moment method is not thermodynamically consistent \cite{note2}.
Actually, Grad's method is not but an approximate scheme for (partially) solving the 
hierarchy of moment equations stemming from the Boltzmann 
equation. 
The point I want to emphasize is that the physical inconsistency 
 of 
the equations for the stress tensor elements derived in 
Refs.~\cite{E02,UGC99} does not lie in the use of Grad's method (even if 
nonlinear terms are neglected), but in the wrong ansatz about the 
stationarity of the flow.
In fact, as said before, the application of Grad's method to the unidirectional flow at uniform temperature for Maxwell molecules gives the same evolution equations for the stress elements as the Boltzmann equation.
Therefore, at least in this instance, Grad's method is free from any thermodynamic inconsistency.

It might be argued that assumptions (a)--(e) [or (a), (c), (d'), and (e)] are used in Refs.~\cite{E02,UGC99} only as a tool to derive \textit{rheological} constitutive equations relating the irreversible part of the stress tensor to the velocity gradients in a \textit{nonlinear} way by means of Grad's method. Such constitutive equations could then be applied to the conservation equations (\ref{1})--(\ref{3}) regardless of whether the flow is steady or not, whether the pressure is uniform or not, etc. However, it is doubtful that a constitutive equation derived from assumptions incompatible with the conservation laws can be acceptable beyond the Navier--Stokes regime, as Fig.~\ref{fig1} illustrates.

Before closing this paper, let me comment on a couple of remarks made in 
Ref.~\cite{E02} which are not directly related to the  
discussion made so far. First, Eu states that ``the shear viscosity is impossible to 
define'' in the absence of transversal velocity gradients and, consequently, 
``it is impossible to measure a shear viscosity without shearing the 
fluid.'' This is misleading. To clarify this point, take the Navier--Stokes 
constitutive equations, namely \cite{CC70},
\beq
P_{ij}=p\delta_{ij}-\eta_{\text{NS}} \left(\frac{\partial u_i}{\partial 
x_j}+\frac{\partial u_j}{\partial x_i}-\frac{2}{3}\nabla\cdot 
\mathbf{u}\,\delta_{ij}\right).
\label{8}
\eeq
In the special case of a unidirectional flow 
$\mathbf{u}=u_x\widehat{\mathbf{x}}$, Eq.~(\ref{8}) yields
\beq
P_{xy}=-\eta_{\text{NS}} \frac{\partial u_x}{\partial y}, \quad 
\frac{1}{2}\left(P_{xx}-P_{yy}\right)=-\eta_{\text{NS}} \frac{\partial u_x}{\partial x}.
\label{9}
\eeq
Thus, to Navier--Stokes order, the response of the shear stress $P_{xy}$ to 
a shear rate $\partial u_x/\partial y$ is the same as the response of the 
normal stress difference $(P_{xx}-P_{yy})/2$ to a longitudinal deformation 
rate $\partial u_x/\partial x$. As a consequence,  the Navier--Stokes shear viscosity can 
be measured from the normal stress difference, even in the absence of 
shearing ($\partial u_x/\partial y=0$).

The second point refers to Eu's claim \cite{E02}  that the 
velocity distribution function $f(\mathbf{r},\mathbf{v};t)$ obeying the 
Boltzmann equation must \textit{always} depend on the three spatial 
coordinates despite the fact that the hydrodynamic variables may depend on 
one space coordinate only, e.g., $n(x,t)$, $\mathbf{u}(x,t)$, $T(x,t)$.
In support of this claim, Eu recalls that ``even if the fluid particle moves 
one dimensionally in its hydrodynamic configuration space, it does not mean 
that the molecules making up the fluid particle and contained in the 
elementary volume of the hydrodynamic configuration space [\ldots] should be 
moving one dimensionally.'' While the quoted sentence is entirely correct, 
Eu's conclusion, namely that one cannot have 
$f(\mathbf{r},\mathbf{v},t)=f(x,\mathbf{v},t)$, does not apply to 
 the Boltzmann velocity distribution function 
$f(\mathbf{r},\mathbf{v},t)$ but to the \textit{microscopic} 
one-body distribution function defined by
\beq
F(\mathbf{r},\mathbf{v},t)=\sum_{i=1}^N \delta(\mathbf{r}-\mathbf{r}_i(t))
\delta(\mathbf{v}-\mathbf{v}_i(t)),
\label{10}
\eeq
where $\{\mathbf{r}_i(t), i=1,\ldots,N\}$ and 
$\{\mathbf{v}_i(t), i=1,\ldots,N\}$ are the sets of positions and velocities 
of the particles of the system at time $t$. Actually, the velocity 
distribution function $f(\mathbf{r},\mathbf{v},t)$ is the \textit{average} 
of $F(\mathbf{r},\mathbf{v},t)$,
\beq
f(\mathbf{r},\mathbf{v},t)=\langle F(\mathbf{r},\mathbf{v},t)\rangle
=\int d\Gamma \, F(\mathbf{r},\mathbf{v},t) \rho(\Gamma),
\label{11}
\eeq
where $\rho(\Gamma)$ is the probability density or ensemble for the initial 
state and the integration is carried out over all the points $\Gamma$ of the 
phase space.
While in a given microscopic realization of the system 
$F(\mathbf{r},\mathbf{v},t)$ is a highly nonuniform function, its 
statistical average $f(\mathbf{r},\mathbf{v},t)$ has a much smoother spatial 
dependence. In particular, it can depend on one coordinate only or it 
can even be uniform (e.g., at equilibrium). Of course, the fact that the 
hydrodynamic fields have a one-dimensional spatial dependence does not 
necessarily mean that the same holds to $f$, but there is nothing wrong if 
one restricts oneself to solutions to the Boltzmann equation with the same 
symmetry properties as the hydrodynamic fields. In fact, the so-called 
normal solutions are those that depend on space and time through a 
\textit{functional} dependence on the hydrodynamic fields \cite{CC70}.

\acknowledgments
 Partial support from the Ministerio de Ciencia y 
Tecnolog\'{\i}a 
 (Spain) and from FEDER through grant No.\ BFM2001-0718 is gratefully acknowledged.


\begin{thebibliography}{41}
\bibitem{note1}
Note that if we perform a rotation on the $y$-$z$ plane and define the new axes $y'=(\gamma_{yx}y+\gamma_{zx}z)/\sqrt{\gamma_{yx}^2+\gamma_{zx}^2}$,
$z'=(\gamma_{zx}y-\gamma_{yx}z)/\sqrt{\gamma_{yx}^2+\gamma_{zx}^2}$, then we have the ``canonical'' form $\partial u_x/\partial y'=\sqrt{\gamma_{yx}^2+\gamma_{zx}^2}$, $\partial u_x/\partial z'=0$.
Thus the true physical independent parameters are the longitudinal deformation rate $\gamma_x$ and the shear rate $\sqrt{\gamma_{yx}^2+\gamma_{zx}^2}$.
\end{thebibliography}

\end{document}